\documentclass[preprint,prb,amsmath,amssymb]{revtex4}

\usepackage{graphicx}
\usepackage{mathptmx}

\usepackage{color}

\begin{document}

\title{Electronic and magnetic properties of the graphene/Eu/Ni(111) hybrid system}

\author{E.~N. Voloshina$^a$ and Yu.~S. Dedkov$^b$}
\affiliation{$^a$\mbox{Humboldt-Universit\"at zu Berlin, Institut f\"ur Chemie, 10099 Berlin, Germany}\\
$^b$\mbox{SPECS Surface Nano Analysis GmbH, Voltastra\ss e~5, 13355 Berlin, Germany}
}


\begin{abstract}
The electronic and magnetic properties of the graphene/Eu/Ni(111) intercalation-like system are studied in the framework of the GGA+U approach with dispersive interactions taken into account. Intercalation of monoatomic Eu layer underneath graphene on Ni(111) leads to the drastic changes of the electronic structure of graphene compared to free-standing graphene as well as graphene/Ni(111). The strong influence of the spin-polarized Eu\,$4f$ states, crossing the graphene-derived $\pi$ states, on magnetic properties of graphene and on spin-filtering properties of the graphene/Eu/Ni(111) trilayer is discussed.   

\vspace{0.44cm}
\hspace{-0.44cm}
\begin{keywords}
\textit{\textit{Key words:} Graphene-metal interfaces; Density functional theory; Electronic structure of gra\-phe\-ne}
\end{keywords}

\vspace{0.44cm}
\hspace{-0.44cm}
Preprint requests to E.N.V.; E-mail: elena.voloshina@hu-berlin.de

\end{abstract}

\maketitle

\section{Introduction}

Graphene, a single layer of carbon atoms arranged in a honeycomb lattice, presently attracts much attention due to its unique physical properties~\cite{Geim:2007a,CastroNeto:2009,Geim:2009}. Many of these properties are defined by the behaviour of the graphene-derived $\pi$ states in the vicinity of the Fermi level ($E_F$), which have a linear dispersion around the $\mathrm{K}$ points of the graphene Brillouin zone (BZ). This also leads to the zero-level density of states (DOS) at $E_F$ attributing graphene to a class of solids called semi-metals. Such properties make graphene an ideal material for the fabrication of different low-dimensional devices which were already made or proposed~\cite{Geim:2009,Novoselov:2011}.

Any application of graphene in the real electron- or spin-transport devices implies the use of the graphene-metal contacts, which can drastically modify electronic as well magnetic properties of graphene. For example, the efficiency of charge- or spin-injection depends on the effective resistance of such graphene/metal interface. Recently, many experimental and theoretical works were devoted to the consideration of the structural, electronic and magnetic properties of the graphene/metal systems. These results were intensively discussed in several review-style articles published in the last five years~\cite{Wintterlin:2009,Batzill:2012}. However, as it can be found, up to now there is no complete understanding of the physical and chemical processes, which may adequately describe or predict properties of a graphene/metal interface~\cite{Voloshina:2012c}.

Recently, in a series of experimental and theoretical works, it was shown that electronic and magnetic properties of the graphene/metal interface can be tailored in different ways that graphene either may behave like free-standing (linear dispersion of the graphene $\pi$ states is preserved) or strongly interacting with substrate (linear dispersion of $\pi$ states is fully destroyed due to the hybridisation with the valence band states of metal). For example, graphene becomes magnetic when it is in contact with ferromagnetic substrate, Ni(111) or Fe/Ni(111)~\cite{Weser:2011}. In this case graphene behaves like ``strongly'' bonded to the substrate due to the strong intermixing of graphene $\pi$ and Ni (or Fe) $3d$ valence band states and the linear dispersion of $\pi$ states is fully destroyed. However, the linear dispersion can be restored and the interaction between graphene and substrate can be weakened upon intercalation of different metals, like Al or noble metals~\cite{Voloshina:2011NJP,Rybkina:2013cn,Shikin:2000,Dedkov:2001,Dedkov:2003,Varykhalov:2008,Varykhalov:2010a}. 

Surprisingly, not many works exist, which deal with the interface between graphene and $4f$ rare earth (RE) metals~\cite{Shikin:2000b,Watcharinyanon:2013fs}, although RE graphite intercalation compound (GICs) were in focus during last two decades~\cite{Molodtsov:1996wd,Molodtsov:1998wk,Shikin:2000vw,Molodtsov:2003jp} because of the interest to the superconductivity of YbC$_6$~\cite{Weller:2005kt,Csanyi:2005ij,Mazin:2005gn,Upton:2010jj}. In order to fill this niche we make a first step in this direction where Eu($6s^24f^7$) single layer is placed in-between graphene and Ni(111). Here we present the systematic DFT studies of the electronic and magnetic properties of the graphene/Eu/Ni(111) intercalation-like system. We found that graphene in this system is strongly $n$-doped and behaviour of its $\pi$ states around the Dirac cone is strongly influenced by the Eu\,$4f$ states demonstrating for spin-up electrons a strong hybridisation between valence band states of graphene and Eu. The linear dispersion of the graphene\,$\pi$ states is conserved for spin-down electrons. Perspectives of application of such interface in possible spin-filtering devices are discussed.

\section{\label{sec:method}Theoretical and computational details}

The DFT calculations were carried out using the projector augmented wave method~\cite{paw}, a plane wave basis set and the generalized gradient approximation as parameterized by Perdew \textit{et al.} (PBE)~\cite{Perdew:1996}, as implemented in the VASP program (v. 5.2.12)~\cite{Kresse:1996}. The plane wave kinetic energy cutoff was set to $500$\,eV. The nuclei and core electrons are represented by frozen cores and PAW potentials, leaving the C $2s2p$, Ni $3d4s$, and Eu $5s5p5f6s$ electrons treated as valence electrons.The strong local Coulomb interaction of the Eu $4f$ electrons is accounted for within the DFT+U approach with the Coulomb parameters of $U=7$\,eV and $J=1$\,eV that are known to be well suited to describe rare earth systems~\cite{Eu-dft+U,gr+Eu}. The long-range van der Waals interactions were accounted for by means of a DFT-D2 approach proposed by Grimme~\cite{Grimme:2004}. The studied system is modelled using a supercell, which has a $(\sqrt{3}\times \sqrt{3})R30^\circ$ overstructure with respect to the unit cell of graphene [Fig.~\ref{structure}(a)] and consists of $53$ atoms:  $13$ layers of Ni atoms ($3$ atoms per layer) with one Eu layer ($1$ atom each) and a graphene sheet ($6$ atoms per layer) adsorbed on both sides of the slab. Metallic slab replicas are separated by about $24$\,\AA\ in the surface normal direction, leading to an effective vacuum region of about $17$\,\AA. In the total energy calculations and during the structural relaxation (the positions of the carbon atoms as well as those of Eu and the top two layers of Ni are optimized) the $k$-meshes for sampling of the supercell Brillouin zone were chosen to be as dense as $24\times 24$ and $12\times 12$, respectively, when folded up to the simple graphene unit cell. 

The scanning tunneling microscopy (STM) images are calculated using the Tersoff-Hamann formalism~\cite{Tersoff:1985}, which states that the tunnelling current in an STM experiment is proportional to the local density of states (LDOS) integrated from the Fermi level to the bias. The STM tip is approximated by an infinitely small point source. The integrated LDOS is calculated as $\bar{\rho}(\mathbf{r},\varepsilon)\propto \int_\varepsilon^{E_F}\rho(\mathbf{r},\varepsilon ')d\varepsilon '$ with $E_F$ the Fermi energy. An STM in constant current mode follows a surface of constant current, which translates into a surface of constant integrated LDOS [$\bar{\rho}(x,y,z,\varepsilon)=C$ with C a real constant]. For each $C$, this construction returns a height $z$ as a function of the position $(x, y)$. This heightmap is then mapped linearly onto a corresponding colour scale.

\section{Results and discussion}

\begin{figure}[h]
\includegraphics[width=9cm]{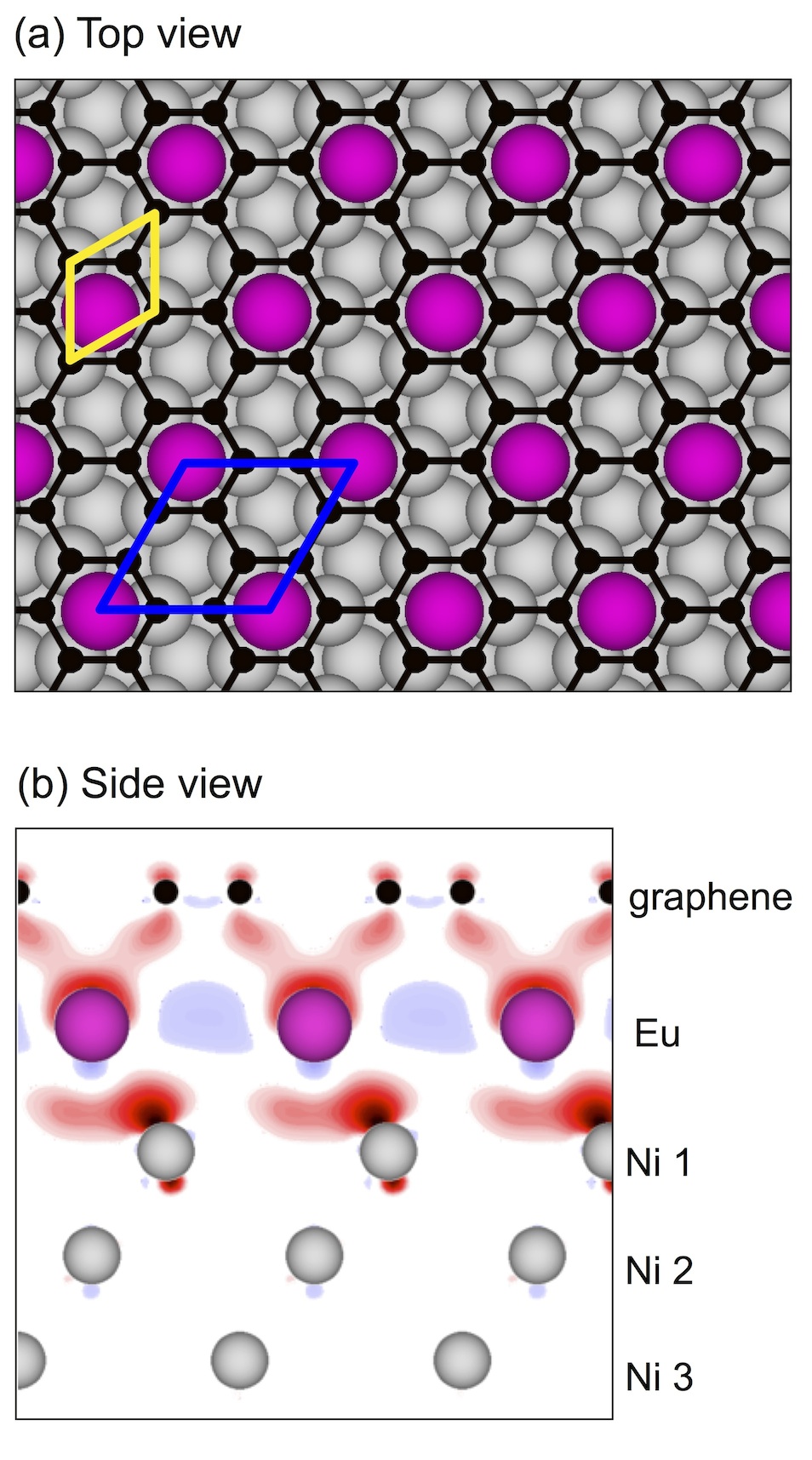}
\caption{\label{structure} (Color online) (a) Top and (b) side views of the graphene/Eu/Ni(111) system with $(\sqrt{3}\times \sqrt{3})R30^\circ$ symmetry. Small, middle, and large color spheres correspond to carbon, Ni, and Eu atoms, respectively. In (a) small and large rhombuses mark unit cells of gr/Ni(111) and gr/Eu/Ni(111), correspondingly. In (b) the structure of the system is overlaid with the calculated difference electron density, $\Delta\rho (r)=\rho_{gr/Eu/Ni(111)}(r)-\rho_{Ni(111)}(r)-\rho_{Eu}(r)-\rho_{gr}(r)$ (red -- accumulation, blue -- depletion of the electron density).}
\end{figure}

The widely accepted structure of graphene/Ni(111) is when carbon atoms are arranged in the so-called \textit{top-fcc} configuration on Ni(111)~\cite{Bertoni:2004}. In this case one of the carbon atoms of the graphene unit cell is placed above the top Ni atom [Ni\,1 in Fig.~\ref{structure}(b)] and the second carbon atom is placed in the $fcc$ hollow site of Ni(111) slab [above Ni\,3 in Fig,~\ref{structure}(b)]. Our earlier and present calculations performed under the same computational settings confirmed this model~\cite{Voloshina:2011NJP,Voloshina:2013cw}.

In case of graphene/Eu/Ni(111), it is assumed that this system has $(\sqrt{3}\times \sqrt{3})R30^\circ$ symmetry with respect to graphene/Ni(111) as this symmetry was found for Eu-GIC~\cite{Molodtsov:1996wd,Shikin:2000vw}. Considering possible crystallographic structures of this intercalation-like system, one can see that Eu atoms below graphene can be placed either in the \textit{FCC} or in the \textit{HCP} hollow sites of the Ni(111) slab or above the interfacial (\textit{TOP}) Ni atom. Our calculations demonstrate that the \textit{HCP} arrangement is significantly (by ca.~$0.25$\,eV) more stable from the energetic point of view. The top and side views of this structure are shown in Fig.~\ref{structure}(a,b), respectively. Here, the distance between graphene layer and underlying Eu is $2.57$\,\AA\ that places the graphene/Eu/Ni(111) between ``strongly'' and ``weakly'' interacting graphene-metal interfaces. 

\begin{figure}[h]
\includegraphics[width=8cm]{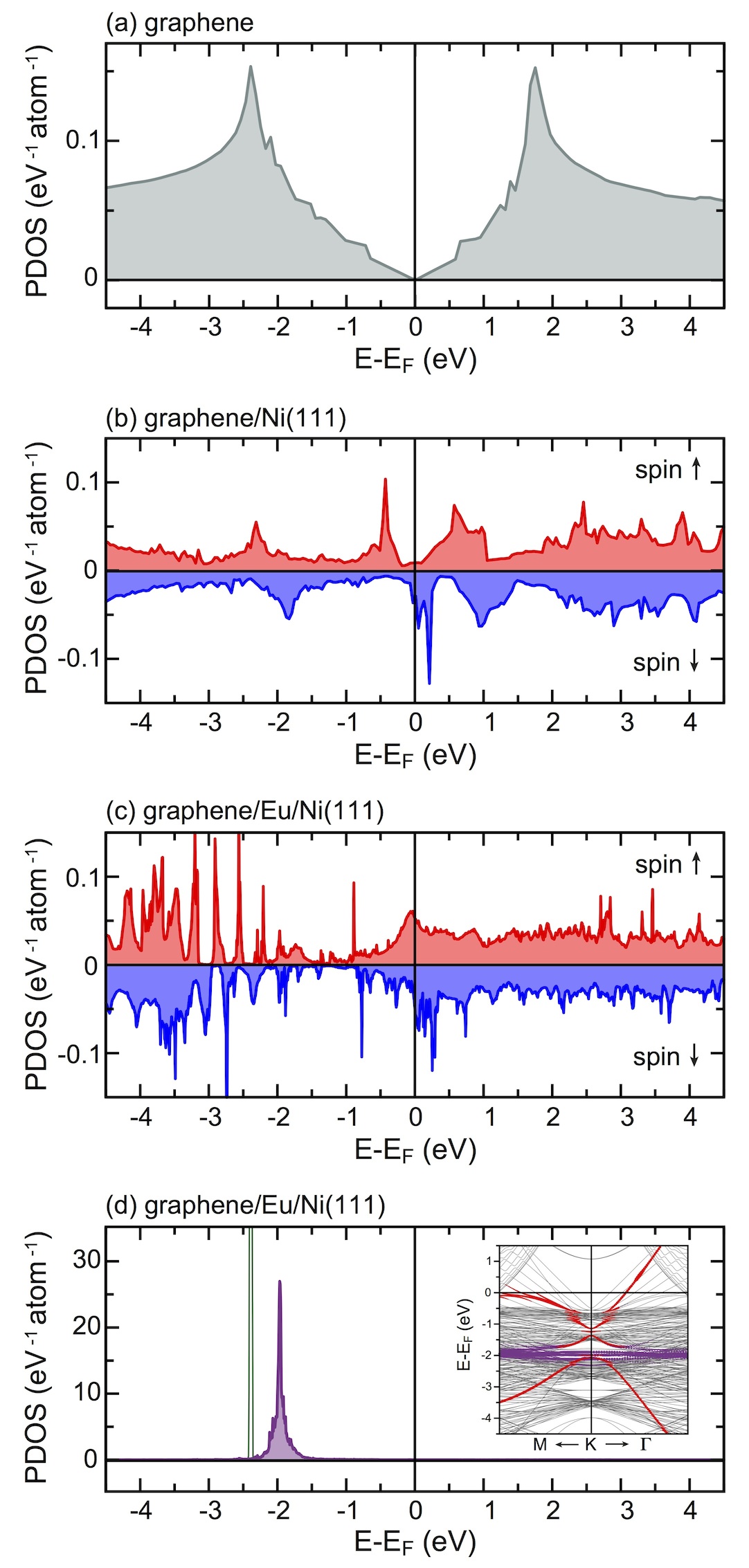}
\caption{\label{dos} (Color online) (a-c) Spin-resolved C-atom projected DOS plots for graphene, graphene/Ni(111), and graphene/Eu/Ni(111), respectively. (d) Eu\,$4f$ partial DOS for graphene/Eu/Ni(111) (filled area) and bulk $bcc$ Eu (solid thin line) presented for comparison. Inset of (d) shows the electronic band structure of graphene/Eu/Ni(111) in the vicinity of the $\mathrm{K}$ point for the spin-up channel. The C\,$p_z$ and Eu\,$4f$ projected bands are shown by red and violet circles, respectively.}
\end{figure}

Intercalation of Eu underneath graphene on Ni(111) leads to the strong modification of the electronic structure of graphene. Fig.~\ref{dos} shows the spin-resolved C-atom projected DOS for (a) free-standing graphene, (b) graphene/Ni(111), and (c) graphene/Eu/Ni(111). The panel (d) shows the respective spin-resolved Eu\,$4f$ partial DOS for the graphene/Eu/Ni(111) system (here only spin-up channel is shown; the spin-down Eu\,$4f$ peak is located at $8.4$\,eV above Fermi level and not shown here). The inset of (d) shows the band structure of the graphene/Eu/Ni(111) system around the $\mathrm{K}$ point for the spin-up channel. The modification of the electronic structure of graphene upon its adsorption on ferromagnetic Ni(111) was discussed earlier in a series of experimental and theoretical works~\cite{Bertoni:2004,Karpan:2008,Khomyakov:2009,Dedkov:2010a,Voloshina:2011NJP}. All these works identify the significant changes in the electronic structure of graphene in this system compared to free-standing graphene: graphene is strongly $n$-doped; there is a strong hybridisation between graphene $\pi$ and Ni\,$3d$ states that leads to the appearance of the induced magnetic moments of carbon atoms in this system~\cite{Weser:2011}; several, so-called, interface Ni\,$3d$--graphene\,$\pi$ states appear in the large band gap between $\pi$ and $\pi^*$ states.

Intercalation of Eu between graphene and Ni leads to the decoupling of the electronic states of graphene from that of the Ni(111) substrate. Graphene in the graphene/Eu/Ni(111) system is $n$-doped due to the partial transfer of the mobile Eu\,$6s$ electrons on the $\pi^*$ states of graphene. The charge transfer from Eu to graphene leads also to the shift of the $4f$ level by $\approx0.5$\,eV to the smaller binding energies [Fig.~\ref{dos}(d)]. As there is no hybridisation between graphene $\pi$ states and $d$ states of substrate, the Dirac cone is restored and it is found at binding energy (BE) of $1.2$\,eV. It is interesting to note that there is a clear hybridisation between graphene $\pi$ and Eu $4f$ valence band states at higher binding energies (compared to the energy of the Dirac point). Since the Eu $4f$ states are strongly spin polarized (due to the contact with ferromagnetic Ni(111) substrate), for the occupied states this hybridization appears only for spin-up electrons in the range of $1.7-2.0$\,eV of BE. This interaction opens a gap of $\approx220$\,meV around the Dirac point of graphene for spin-up channel. At the same time the linear dispersion of graphene $\pi$ states for spin down electrons remains intact with a band gap of $\approx130$\,meV around Dirac point (now shown here). This effect correlates with the conclusion made recently in Ref.~\cite{Voloshina:2012c}, where authors claim that only broken sublattice symmetry for two carbon atoms in the unit cell of graphene accompanied with the hybridisation of the graphene $\pi$ and substrate valence band state might lead to the opening of the larger gap at the Dirac point. This effect is clearly confirmed here for the same carbon atoms, but for two different spin channels, where for one of them the hybridisation between graphene $\pi$ states and Eu $4f$ exists. The effect of different gap width for different spin channel can be used for the fabrication of spin-filtering device where the position of the gap with respect to the Fermi level can be tuned by the external electric potential. The details of such analysis will be presented in our future publications.

The effect of hybridisation between graphene $\pi$ and Eu\,$4f$ valence band states is clearly visible in the difference electron density map shown in Fig.~\ref{structure}(b). Here the formation of the hybrid states involving graphene $\pi$ and Eu\,$4f$ electrons is visible [due to the symmetry of the system, the $4f$ orbitals having projections on the $z$-axis (excluding $4f_{z^3}$) are involved in the formation of these states].

As mentioned earlier, the Eu layer is strongly spin polarised due to the contact with the underlying Ni(111) substrate. The effect of hybridization between $\pi$ and $4f$ states (as well as proximity effect~\cite{Manna:2013bp}) might lead to the appearance of the magnetic moment on carbon atoms. In fact, the calculated magnetic moment of carbon atoms is $0.006\mu_B$. This value obtained from calculations is surprisingly small because the magnetic exchange splitting of the valence band states extracted from DOS plots, which is in the range of $0.3-0.4$\,eV for different states [Fig.~\ref{dos}(c)], can give a large value of magnetic moment of carbon atoms. The experimental verification by means of spin-resolved photoemission or x-ray magnetic circular dichroism is necessary~\cite{Weser:2011}.  

\begin{figure}[h]
\includegraphics[width=10cm]{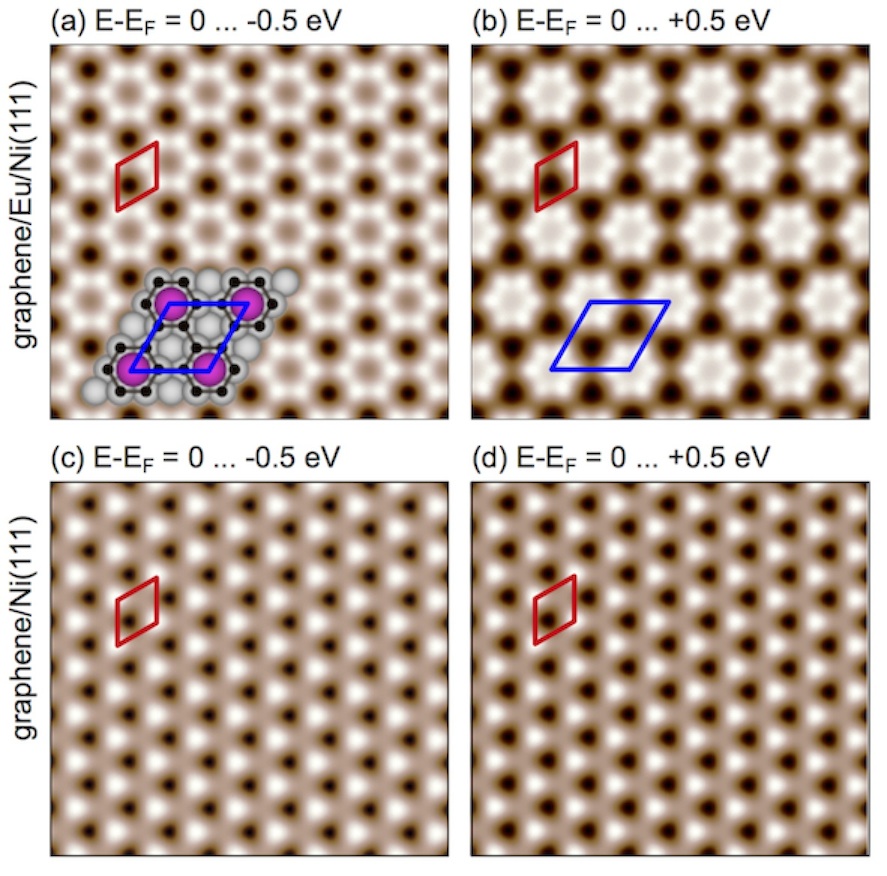}
\caption{\label{stm} (Color online) Simulated STM images of the graphene/Eu/Ni(111) (a,b) and graphene/Ni(111) (c,d) systems for occupied (left) and unoccupied (right) states, respectively. In (a) the STM picture is overlaid with the crystallographic structure of the graphene/Eu/Ni(111) system. Small and large rhombuses in all images correspond to the unit cell of graphene/Ni(111) and graphene/Eu/Ni(111), respectively.}
\end{figure}

The resulting distribution of the electronic states in the real and energy space for the gra\-phe\-ne/Eu/Ni(111) system was used for the simulation of the STM images in order to make comparison between theoretical results and future structural experiments. The resulting pictures are presented in Fig.~\ref{stm}(a,b) where integration energy range is equal to the difference between $E_F$ and binding energy identical to the used bias voltage. First of all, the STM images are clearly different for the bias voltages corresponding to the tunnelling from occupied states [Fig.~\ref{stm}(a)] and on the unoccupied states [Fig.~\ref{stm}(b)] of the system. These results also demonstrate that the sublattice symmetry for two carbon atoms is broken for every carbon ring surrounding either Eu atom or $hcp$ hollow site of the Ni(111) slab. In this case carbon atoms in the ring are imaged with the different topographic or current contrast. In order to perform more careful comparison between theory and experiment the systematic tunnelling microscopy/spectroscopy measurements are necessary. For comparison, the simulated STM images of graphene/Ni(111) are presented in Fig.~\ref{stm}(c,d)~\cite{Dedkov:2010a}.

\section{Conclusion}

We performed DFT studies (GGA+U including long-range dispersive corrections) of the intercalation-like graphene/Eu/Ni(111) system. We found that intercalation of monolayer of Eu leads to the decoupling of the graphene electronic states from those of the substrate. Graphene is strongly $n$-doped in this system and hybridization between graphene $\pi$ states and Eu\,$4f$ states is found in the energy range below Dirac point. This hybridization leads to the lifting of the degeneracy between two carbon atoms in the graphene unit cell that can be detected in the microscopic and spectroscopic experiments, which are proposed and discussed. The different band gaps for the spin-up and spin-down channels in the electronic structure of graphene open perspectives for the application of this system in the future spintronic devices where spin-transport properties of graphene can be tuned by the external electric field. 

\section*{Acknowledgements}

\noindent The support from the German Research Foundation (DFG) through the grant VO1711/3-1 within the Priority Program 1459 ``Graphene'' is appreciated. The High Performance Computing Network of Northern Germany (HLRN) is acknowledged for computer time.

\clearpage
\begin{table*}
\caption{\label{table_structures} Relative energies (i.e. the energy difference between the energies calculated for the different slab models and the energy calculated for the \textit{top-fcc\_HCP} arrangement) and the corresponding graphene-Eu and Eu-Ni distances of different structures of graphene/Eu/Ni(111) where Eu atoms are placed in \textit{TOP}, \textit{FCC}, or \textit{HCP} positions above Ni(111). The graphene-Ni distance for graphene/Ni(111) is presented for comparison.}
\vspace{0.35cm}
\begin{tabular}{|l|c|c|c|c|}
\hline
System  &graphene/Ni(111)&\multicolumn{3}{c|}{graphene/Eu/Ni(111)}\\
\cline{2-5}
        &\textit{top-fcc}&\textit{top-fcc\_TOP}&\textit{top-fcc\_FCC}&\textit{top-fcc\_HCP}\\
\hline
Rel. energy (meV) &      &$+251$&$+252$&$0$\\
$d$ (gr-Ni) (\AA)&$2.08$& &&\\
$d$ (gr-Eu) (\AA)&      &$2.74$&$2.75$&$2.57$\\
$d$ (Eu-Ni) (\AA)&      &$2.41$&$2.42$&$2.45$\\
\hline
\end{tabular}
\end{table*}

\clearpage

\clearpage

\clearpage


\begin{thebibliography}{99}
\bibitem{Geim:2007a}
A. K. Geim and K. S. Novoselov, Nature Materials \textbf{6}, 183 (2007).
\bibitem{CastroNeto:2009}
A. Castro Neto, F. Guinea, N. Peres, K. Novoselov, and A. Geim, Rev. Mod. Phys. \textbf{81}, 109 (2009).
\bibitem{Geim:2009}
A. Geim, Science \textbf{324}, 1530 (2009).
\bibitem{Novoselov:2011}
K. Novoselov, Rev. Mod. Phys. \textbf{83}, 837 (2011).
\bibitem{Wintterlin:2009}%
J. Wintterlin and M. L. Bocquet, Surf. Sci. \textbf{603}, 1841 (2009).
\bibitem{Batzill:2012}
M. Batzill, Surf. Sci. Rep. \textbf{67}, 83 (2012).
\bibitem{Voloshina:2012c}
E. Voloshina and Yu. Dedkov, Phys. Chem. Chem. Phys. \textbf{14}, 13502 (2012).
\bibitem{Weser:2011}
M. Weser, E. N. Voloshina, K. Horn, and Yu. S. Dedkov, Phys. Chem. Chem. Phys. \textbf{13}, 7534 (2011).
\bibitem{Voloshina:2011NJP}
E. N. Voloshina, A. Generalov, M. Weser, S. B\"ottcher, K. Horn, and Yu. S. Dedkov, New J. Phys. \textbf{13}, 113028 (2011).
\bibitem{Rybkina:2013cn}
A. A. Rybkina, A. G. Rybkin, A. V. Fedorov, D. Y. Usachov, M. E. Yachmenev, D. E. Marchenko, O. Y. Vilkov, A. V. Nelyubov, V. K. Adamchuk, and A. M. Shikin, Surf. Sci. \textbf{609}, 7 (2013).
\bibitem{Shikin:2000}
A. Shikin, G. Prudnikova, V. Adamchuk, F. Moresco, and K. Rieder, Phys. Rev. B \textbf{62}, 13202 (2000).
\bibitem{Dedkov:2001}
Yu. S. Dedkov, A. M. Shikin, V. K. Adamchuk, S. L. Molodtsov, C. Laubschat, A. Bauer, and G. Kaindl, Phys. Rev. B \textbf{64}, 035405 (2001).
\bibitem{Dedkov:2003}
Yu. S. Dedkov, M. Poygin, D. Vyalikh, A. Starodubov, A. M. Shikin, and V. K. Adamchuk, arXiv:cond-mat/0304575v1 [cond-mat.mtrl-sci] (2003).
\bibitem{Varykhalov:2008}
A. Varykhalov, J. Sanchez-Barriga, A. M. Shikin, C. Biswas, E. Vescovo, A. Rybkin, D. Marchenko, and O. Rader, Phys. Rev. Lett. \textbf{101}, 157601 (2008).
\bibitem{Varykhalov:2010a}
A. Varykhalov, M. Scholz, T. Kim, and O. Rader, Phys. Rev. B \textbf{82} 121101(R) (2010).
\bibitem{Shikin:2000b}
A. Shikin, M. Poigin, Y. Dedkov, S. Molodtsov, and V. Adamchuk, Phys. Solid State \textbf{42}, 1170 (2000).
\bibitem{Watcharinyanon:2013fs}
S. Watcharinyanon, L. I Johansson, C. Xia, J. Ingo Flege, A. Meyer, J. Falta, and C. Virojanadara, Graphene \textbf{02}, 66 (2013).
\bibitem{Molodtsov:1996wd}
S. Molodtsov, C. Laubschat, M. Richter, T. Gantz, and A. Shikin, Phys. Rev., B Condens. Matter \textbf{53}, 16621 (1996).
\bibitem{Molodtsov:1998wk}
S. L. Molodtsov, Journal of Electron Spectroscopy and Related Phenomena \textbf{96}, 157 (1998).
\bibitem{Shikin:2000vw}
A. M. Shikin, V. K. Adamchuk, S. Siebentritt, K. H. Rieder, S. L. Molodtsov, and C. Laubschat, Phys. Rev. B \textbf{61}, 7752 (2000).
\bibitem{Molodtsov:2003jp}
S. Molodtsov, F. Schiller, S. Danzenb\"acher, M. Richter, J. Avila, C. Laubschat, and M. Asensio, Phys. Rev. B \textbf{67} (2003).
\bibitem{Weller:2005kt}
T. E. Weller, M. Ellerby, S. S. Saxena, R. P. Smith, and N. T. Skipper, Nature Physics \textbf{1}, 39 (2005).
\bibitem{Csanyi:2005ij}
G. Cs\'anyi, P. B. Littlewood, A. H. Nevidomskyy, C. J. Pickard, and B. D. Simons, Nature Physics \textbf{1}, 42 (2005).
\bibitem{Mazin:2005gn}
I. Mazin and S. Molodtsov, Phys. Rev. B \textbf{72}, 172504 (2005).
\bibitem{Upton:2010jj}
M. H. Upton, T. R. Forrest, A. C. Walters, C. A. Howard, M. Ellerby, A. H. Said, and D. F. McMorrow, Phys. Rev. B \textbf{82}, 134515 (2010).
\bibitem{paw}
P.Bl\"ochl, Phys. Rev. B \textbf{50}, 17953 (1994).
\bibitem{Perdew:1996}
J. Perdew, K. Burke, and M. Ernzerhof, Phys. Rev. Lett. \textbf{77}, 3865 (1996).
\bibitem{Kresse:1996}
G. Kresse and J. Furthmuller, Phys. Rev. B \textbf{54}, 11169 (1996).
\bibitem{Eu-dft+U}
V. I. Anisimov, F. Aryasetiawan, A. I. Lichtenstein, J. Phys.: Condens. Matter \textbf{9}, 767 (1997).
\bibitem{gr+Eu}
S. Schumacher, T. O. Wehling, P. Lazi\'c, S. Runte, D. F. F\"orster, C. Busse, M. Petrovi\c{c}, M. Kralj, S. Bl\"ugel, N. Atodiresei, V. Caciuc, and T. Michely, Nano Lett. \textbf{13}, 5013 (2013).
\bibitem{Grimme:2004}
S. Grimme, J. Comput. Chem. \textbf{25}, 1463 (2004); S. Grimme, J. Comput. Chem. \textbf{27}, 1787 (2006).
\bibitem{Tersoff:1985}
J. Tersoff and D. R. Hamann, Phys. Rev. B \textbf{31}, 805 (1985).
\bibitem{Bertoni:2004}
G. Bertoni, L. Calmels, A. Altibelli, and V. Serin, Phys. Rev. B \textbf{71}, 075402 (2004).
\bibitem{Voloshina:2013cw}
E. Voloshina, R. Ovcharenko, A. Shulakov, and Yu. S. Dedkov, J. Chem. Phys. \textbf{138}, 154706 (2013).
\bibitem{Karpan:2008}
V. M. Karpan, P. A. Khomyakov, A. A. Starikov, G. Giovannetti, M. Zwierzycki, M. Talanana, G. Brocks, J. v. d. Brink, and P. J. Kelly, Phys. Rev. B \textbf{78}, 195419 (2008).
\bibitem{Khomyakov:2009}
P. A. Khomyakov, G. Giovannetti, P. C. Rusu, G. Brocks, J. v. d. Brink, and P. J. Kelly, Phys. Rev. B \textbf{79}, 195425 (2009).
\bibitem{Dedkov:2010a}
Yu. S. Dedkov and M. Fonin, New J. Phys. \textbf{12}, 125004 (2010).
\bibitem{Manna:2013bp}
P. K. Manna and S. M. Yusuf, Physics Reports (2013), http://dx.doi.org/10.1016/j.physrep.2013.10.002.

\end{thebibliography}
\end{document}